\documentclass[12pt]{article}
\usepackage{epsf}
\usepackage{epsfig}
\usepackage{wrapfig}
\usepackage{amsmath}
\usepackage{amssymb}
\usepackage{eufrak}
\usepackage{graphicx}
\begin{document}

\begin{center}
\section*{Collapse of Bose component in Bose-Fermi mixture with attraction between components}

\vskip 5mm S.-T. Chui$^{1}$, V. N. Ryzhov$^{2}$ and E. E.
Tareyeva$^{2}$ \vskip 5mm

{\small (1) {\it Bartol Research Institute, University of
Delaware, Newark, DE 19716}
\\
(2) {\it Institute for High Pressure Physics, Russian Academy of
Sciences, 142 190 Troitsk, Moscow region, Russia}

}
\end{center}
\vskip 5mm

\begin{abstract}
An effective Hamiltonian for the Bose system in the mixture of
ultracold atomic clouds of bosons and fermions is obtained by
integrating out the Fermi degrees of freedom. An instability of
the Bose system is found in the case of attractive interaction
between components in good agreement with the experiment on the
bosonic $^{87}$Rb and fermionic $^{40}$K mixture.
\end{abstract}
\vskip 8mm

The seminal paper by N.N. Bogoliubov "To the theory of
superffluidity", published in 1947, had a great influence on the
development of the modern theory of quantum many-body systems. It
is cited in all textbooks devoted to this topic. However, for a
long time the real experimental systems which could be
quantitatively described by the model of dilute Bose gas developed
in this article were absent. Only about 50 years later, the
Bose-Einstein condensation was achieved in the dilute ultracold
gas of magnetically trapped alkali atoms. Since the first
realization of Bose-Einstein condensation in ultracold atomic gas
clouds \cite{[1],[3]}, studies in this direction have yielded
unprecedented insight into the quantum statistical properties of
matter. Besides the studies using the bosonic atoms, growing
interest is focused on the cooling of fermionic atoms to a
temperature regime where quantum effects dominate the properties
of the gas \cite{sci99}. This interest is mainly motivated by the
quest for the Bardeen-Cooper-Schrieffer (BCS) transition in
ultarcold atomic Fermi gasses \cite{nat03_1}.

Strong {\it s}-wave interactions that facilitate evaporative
cooling of bosons are absent among spin-polarized fermions due to
the exclusion Pauli principle. So the fermions are cooled to
degeneracy through the mediation of fermions in another spin state
\cite{sci99,nat03_1} or via a buffer gas of bosons \cite{sci02bf}
(sympathetic cooling). The Bose gas, which can be cooled
evaporatively, is used as a coolant, the fermionic system being in
thermal equilibrium with the cold Bose gas through boson-fermion
interaction in the region of overlapping of the systems.

However, the physical properties of Bose-Fermi mixtures are
interesting in their own rights and are the subject of intensive
investigations including the analysis of ground state properties,
stability, effective Fermi-Fermi interaction mediated by the
bosons, and new quantum phases in an optical lattices. Several
successful attempts to trap and cool mixtures of bosons and
fermions were reported. Quantum degeneracy was first reached with
mixtures of bosonic $^7$Li and fermionic $^6$Li atoms. Later,
experiments to cool mixtures of $^{23}$Na and $^6$Li, as well as
$^{87}$Rb and $^{40}$K \cite{sci02bf}, to ultralow temperatures
succeeded.

In this article we study the instability and collapses of the
trapped boson-fermion mixture due to the boson-fermion attractive
interaction, using the effective Hamiltonian for the Bose system
\cite{ChRyz04,CTR04}. We analyze quantitatively properties of the
$^{87}$Rb and $^{40}$K mixture  with an attractive interaction
between bosons and fermions recently studied by Modugno and
co-workers \cite{sci02bf}. They found that as the number of bosons
is increased there is an instability value $N_{Bc}$ at which a
discontinues leakage of of the bosons and fermions occurs, and
collapse of boson and fermion clouds is observed. Using the
experimental parameters we estimated the instability boson number
$N_{Bc}$ for the collapse transition as a function of the fermion
number and temperature and found a good agreement with
experimental results.

First of all we briefly discuss the effective boson Hamiltonian
\cite{ChRyz04,CTR04}.  Our starting point is the
functional-integral representation of the grand-canonical
partition function of the Bose-Fermi mixture. It has the form
\cite{popov1}:
\begin{eqnarray}
Z&=&\int
D[\phi^*]D[\phi]D[\psi^*]D[\psi]\exp\left\{-\frac{1}{\hbar}\left(
S_B(\phi^*,\phi)+\right.\right.\\
\nonumber
&+&\left.\left.S_F(\psi^*,\psi)+S_{int}(\phi^*,\phi,\psi^*,\psi)\right)\right\}.
\label{1}
\end{eqnarray}
and consists of an integration over a complex field
$\phi(\tau,{\bf r})$, which is periodic on the imaginary-time
interval $[0,\hbar\beta]$, and over the Grassmann field
$\psi(\tau,{\bf r})$, which is antiperiodic on this interval.
Therefore, $\phi(\tau,{\bf r})$ describes the Bose component of
the mixture, whereas $\psi(\tau,{\bf r})$ corresponds to the Fermi
component. The term describing the Bose gas has the form:
\begin{eqnarray}
S_B(\phi^*,\phi)&=&\int_0^{\hbar\beta}d\tau\int d{\bf
r}\left\{\phi^*(\tau,{\bf r})\left(\hbar \frac{\partial}{\partial
\tau}-\frac{\hbar^2\nabla^2}{2 m_B} + \right.\right.\nonumber\\
 &+&V_B({\bf r})-\left.\left.\mu_B\right)\phi(\tau,{\bf
r})+ \frac{g_B}{2}|\phi(\tau,{\bf r})|^4\right\}. \label{2}
\end{eqnarray}
Because the Pauli principle forbids $s$-wave scattering between
fermionic atoms in the same hyperfine state, the Fermi-gas term
can be written in the form:
\begin{eqnarray}
S_F(\psi^*,\psi)&=&\int_0^{\hbar\beta}d\tau\int d{\bf
r}\left\{\psi^*(\tau,{\bf r})\left(\hbar \frac{\partial}{\partial
\tau}-\frac{\hbar^2\nabla^2}{2 m_F}+ \right.\right.\nonumber\\
&+&\left.\left.V_F({\bf r}) -\mu_F\right)\psi(\tau,{\bf r})
\right\}. \label{3}
\end{eqnarray}
The term describing the interaction between the two components of
the Fermi-Bose mixture is:
\begin{equation}
S_{int}(\phi^*,\phi,\psi^*,\psi)=g_{BF}\int_0^{\hbar\beta}d\tau\int
d{\bf r} |\psi(\tau,{\bf r})|^2|\phi(\tau,{\bf r})|^2, \label{4}
\end{equation}
where $g_B=4\pi \hbar^2a_B/m_B$ and $g_{BF}=2\pi
\hbar^2a_{BF}/m_I$, $m_I=m_B m_F/(m_B+m_F)$, $m_B$ and $m_F$ are
the masses of bosonic and fermionic atoms respectively, $a_B$ and
$a_{BF}$ are the $s$ wave scattering lengths of boson-boson and
boson-fermion interactions.

Integral over Fermi fields is Gassian, we can calculate this
integral and obtain the partition function of the Fermi system as
a functional of Bose field $\phi(\tau, {\bf r})$. In the
semiclassical Thomas-Fermi approximation one has
\cite{ChRyz04,CTR04}:
\begin{eqnarray}
S_{eff}&=&\int_0^{\hbar\beta}d\tau d{\bf r}f_{eff}(|\phi(\tau,{\bf
r})|),\label{18}\\
f_{eff}&=&-\frac{3}{2}\kappa\beta^{-1}\int_0^\infty\sqrt{\epsilon}
d\epsilon\ln\left(1+e^{\beta
(\tilde{\mu}-\epsilon)}\right)=\nonumber\\
&=&-\kappa\int_0^\infty\frac{\epsilon^{3/2}d\epsilon}{1+e^{\beta(\epsilon-\tilde{\mu})}},
\label{20}
\end{eqnarray}
where $\epsilon=p^2/2m_F$, $\tilde{\mu}=\mu_F-V_F({\bf
r})-g_{BF}|\phi(\tau,{\bf r})|^2$ and
$\kappa=2^{1/2}m_F^{3/2}/(3\pi^2\hbar^3)$.

So we can write the effective bosonic Hamiltonian in the form:
\begin{eqnarray}
H_{eff}&=&\int d{\bf
r}\left\{\frac{\hbar^2}{2m_B}|\nabla\phi|^2+(V_B({\bf
r})-\mu_B)|\phi|^2 +\right.\nonumber
\\ &+&\left.\frac{g_B}{2}|\phi|^4+f_{eff}(|\phi|)\right\}. \label{21}
\end{eqnarray}
The first three terms in (\ref{21}) have the conventional
Gross-Pitaevskii form, and the last term is a result of
boson-fermion interaction. In low temperature limit
$\tilde{\mu}/(k_BT)\gg 1$ one can write $f_{eff}(|\phi|)$ in the
form:
\begin{equation}
f_{eff}(|\phi|)=-\frac{2}{5}\kappa\tilde{\mu}^{5/2}-\frac{\pi^2}{4}\kappa(k_BT)^2
\tilde{\mu}^{1/2}. \label{25}
\end{equation}

As usual, $\mu_F$ can be determined from the equation
\begin{equation}
N_F=\int d{\bf r}n_F({\bf r}), \label{33}
\end{equation}
where
\begin{equation}
n_F({\bf r})=
\frac{3}{2}\kappa\int_0^\infty\frac{\sqrt{\epsilon}d\epsilon}{1+e^{\beta
(\epsilon-\tilde{\mu})}}. \label{31}
\end{equation}

Let us consider now the $^{87}$Rb and $^{40}$K mixture  with an
attractive interaction between bosons and fermions \cite{sci02bf}.
The parameters of the system are the following: $a_B=5.25\,\, nm$,
$a_{BF}=-21.7^{+4.3}_{-4.8}\,\, nm$. K and Rb atoms were prepared
in the doubly polarized states $|F=9/2, m_F=9/2>$ and $|2,2>$,
respectively. The magnetic potential had an elongated symmetry,
with harmonic oscillation frequencies for Rb atoms
$\omega_{B,r}=\omega_B=2\pi\times 215 Hz$ and $
\omega_{B,z}=\lambda\omega_B=2\pi\times 16.3 Hz,
\omega_F=\sqrt{m_B/m_F}\omega_B\approx 1.47 \omega_B$, so that
$m_B\omega_B^2/2=m_F\omega_F^2/2=V_0$.  The collapse was found for
the following critical numbers of bosons and fermions:
$N_{Bc}\approx 10^{5}; N_K\approx 2\times 10^{4}$.

At the zero temperature limit, expanding $f_{eff}(|\phi|)$ up to
the third order in $g_{BF}$ we obtain the effective Hamiltonian in
the form:
\begin{eqnarray}
H_{eff}&=&\int d{\bf
r}\left\{\frac{\hbar^2}{2m_B}|\nabla\phi|^2+(V_{eff}({\bf
r})-\mu_B)|\phi|^2 +\right. \nonumber\\
&+&\left.\frac{g_{eff}}{2}|\phi|^4+\frac{\kappa}{8\mu_F^{1/2}}g_{BF}^3|\phi|^6\right\}.
\label{he}
\end{eqnarray}
where
\begin{eqnarray}
V_{eff}({\bf r})&=&\left(1-\frac{3}{2}\kappa\mu_F^{1/2}
g_{BF}\right)\frac{1}{2}m_B\omega^2_B\left(\rho^2+\lambda^2
z^2\right),\label{p1}\\
g_{eff}&=&g_B-\frac{3}{2}\kappa\mu_F^{1/2}g_{BF}^2,\label{p2}
\end{eqnarray}
and  $\rho^2=x^2+y^2$.

In principle, one can study the properties of a Bose-Fermi mixture
with the help of $f_{eff}$ (\ref{25}) without any expansion.
However, the form of the Hamiltonian (\ref{he}) gives the
possibility to get a clear insight into the physics of the
influence of the Fermi system on the Bose one (see discussion
below). It may be easily verified that the expansion of the
function $f(x)=(1+x)^{5/2}$ (see Eq. (\ref{25})) up to the third
order in $x$ gives a reasonably good approximation for $f(x)$ even
for rather large values of $x$, in contrast with the higher order
expansions, so one can safely use Eq. (\ref{he}) as a starting
point for the investigation of the properties of the Bose
subsystem.

In derivation of Eqs. (\ref{he}-\ref{p2}) we also use the fact
that due to the Pauli principle (quantum pressure) the radius of
the Bose condensate is much less than the radius of the Fermi
cloud $R_F\approx \sqrt{\mu_F/V_0}$, so one can use an expansions
in powers of $V_F({\bf r})/\mu_F$.

From Eq. (\ref{p1}) one can see that the interaction with Fermi
gas leads to modification of the trapping potential. For the
attractive fermion-boson interaction the system should behave as
if it was confined in a magnetic trapping potential with larger
frequencies than the actual ones, in agreement with experiment
\cite{sci02bf}. Boson-fermion interaction also induces the
additional attraction between Bose atoms which does not depend on
the sign of $g_{BF}$.

The last term in $H_{eff}$ (\ref{he}) corresponds to the
three-particle \emph{elastic} collisions induced by the
boson-fermion interaction. In contrast with \emph{inelastic}
3-body collisions which result in the recombination and removing
particles from the system \cite{kagan2}, this term for $g_{BF}<0$
leads to increase of the gas density in the center of the trap in
order to lower the total energy. The positive zero point energy
and boson-boson repulsion energy (the first two terms in Eq.
(\ref{he})) stabilize the system. However, if the central density
grows too much, the kinetic energy and boson-boson repulsion are
no longer able to prevent the collapse of the gas. Likewise the
case of Bose condensate with attraction (see, for example,
\cite{kagan2}), the collapse is expected to occur when the number
of particles in the condensate exceeds the critical value
$N_{Bc}$.

The critical number $N_{Bc}$ can be calculated using the
well-known ansatz for the Bosonic wave function:
\begin{equation}
\phi({\bf
r})=\left(\frac{N_B\lambda}{w^3a^3\pi^{3/2}}\right)^{1/2}
\exp\left(-\frac{\left(\rho^2+\lambda^2
z^2\right)}{2w^2a^2}\right), \label{38}
\end{equation}
where $w$ is a dimensionless variational parameter which fixes the
width of the condensate and $a=\sqrt{\hbar/m_B\omega_B}$.

In this case the variational energy $E_B$ has the form:
\begin{equation}
\frac{E_B}{N_B\hbar\omega_B}=\frac{2+\lambda}{4}\frac{1}{w^2}+b
w^2+\frac{c_1N_B}{w^3}+\frac{c_2N_B^2}{w^6},\label{var}
\end{equation}
\begin{eqnarray*}
b&=&\frac{3}{4}\left(1-\frac{3}{2}\kappa\mu_F^{1/2}g_{BF}\right),\\
c_1&=&\frac{1}{2}\left(g_B-\frac{3}{2}\kappa\mu_F^{1/2}g_{BF}^2\right)
\frac{\lambda}{(2\pi)^{3/2}\hbar\omega_Ba^3},\\
c_2&=&\frac{\kappa}{8\mu_F^{1/2}}g_{BF}^3\frac{\lambda^2}{3^{3/2}
\pi^3\hbar\omega_Ba^6}.
\end{eqnarray*}
This energy is plotted in Fig.1 as a function of $w$ for several
values of $N_B$. It is seen that when $N_B<N_{Bc}$ there is a
local minimum of $E_B$ which correspond to a metastable state of
the system. This minimum arises due the competition between the
positive first three terms in Eq. (\ref{var}) and negative fourth
term. The local minimum disappears when the number of bosons $N_B$
exceeds the critical value which can be calculated by requiring
that the first and second derivatives of $E_B$ vanish at the
critical point. In this case the behavior of $E_B$ is mainly
determined by the second and fourth terms in Eq. (\ref{var}). For
$N_K= 2\times 10^{4}$ and $a_{BF}=-19.44\, nm$ we obtain
$N_{Bc}\approx 9\times10^4$ in a good agreement with the
experiment \cite{sci02bf}. It is interesting to note that the
critical number of Bose atoms in Bose-Fermi mixture is about two
orders larger than the critical number for the condensate with a
purely attractive interaction. For example, in the experiments
with trapped $^7$Li \cite{[3]} it was found that the critical
number of bosons is about $1000$.

\begin{figure}
\begin{center}
\includegraphics[width=8cm]{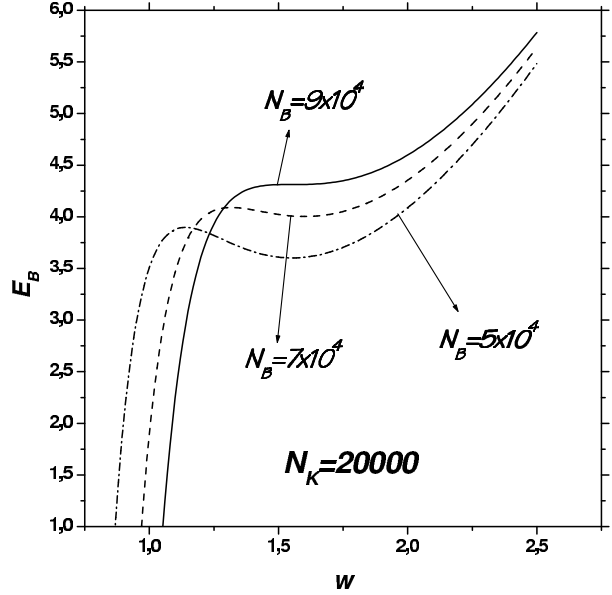}
\caption{Variational energy $\frac{E_B}{N_B\hbar\omega_B}$ as a
function of $w$ for various numbers of bosons.}
\end{center}
\end{figure}

Upon increasing the number of fermions, the repulsion between
bosons decreases leading to the collapse for the smaller numbers
of the bosonic atoms. It should be also noted that an increase of
the temperature results in a decrease of the local density of
fermions and reduces the interaction energy between Bose and Fermi
systems increasing the critical number of bosons.

Finally, we make a short remark on the nature of the collapse
transition. In this article we found the instability point of the
Bose-Fermi mixture with attractive interaction between components.
A strong rise of density of bosons and fermions (in the collapsing
condensate enhances intrinsic inelastic processes, in particular,
the recombination in 3-body interatomic collisions, as is the case
for the well-known $^7$Li condensates \cite{kagan2}. However,
recently M. Yu. Kagan and coworkers suggested the new microscopic
mechanism of removing atoms from the system which is specific for
the Bose-Fermi mixtures with attraction between components and is
based on the formation of the boson-fermion bound states
\cite{MKag}. It seems that the description of the evolution of the
collapsing condensate should include both these mechanisms.

VNR and EET are grateful to N.M. Plakida for the interest to this
work. VNR acknowledges valuable discussions with M. Yu. Kagan. The
work was supported in part by the Russian Foundation for Basic
Research (Grants No 02-02-16622 and No 02-02-16621). STC is partly
supported by a grant from NASA.

\end{document}